\documentclass[twocolumn,showpacs,preprintnumbers,amsmath,amssymb,floatfix]{revtex4}

\usepackage{graphicx}
\usepackage{dcolumn}
\usepackage{bm}

\newcommand{\beq}{\begin{equation}}
\newcommand{\eeq}{\end{equation}}
\newcommand{\bey}{\begin{eqnarray}}
\newcommand{\eey}{\end{eqnarray}}

\begin{document}

\preprint{}

\title{A new class of stable $(2+1)$ dimensional thin shell wormhole}

\author{Farook Rahaman}
\email{rahaman@iucaa.ernet.in} \affiliation{Department of
Mathematics, Jadavpur University, Kolkata 700 032, West Bengal,
India.}
\author{ A. Banerjee}
 \email{ayan\_7575@yahoo.co.in}
\affiliation{Department of Mathematics, Birati Vidyalaya Boys'
,Birati ,Kolkata-700051, India}
\author{I. Radinschi}
\email{radinschi@yahoo.com} \affiliation{ Department of Physics,
"Gh. Asachi" Technical University, Iasi, 700050, Romania.}

\date{\today}

\begin{abstract}
Recently, Ba\~{n}ados, Teitelboim and Zanelli (BTZ) \cite{BTZ1992}
has discovered an explicit vacuum solution of (2+1)-dimensional
gravity with negative cosmological constant. It has been argued
that the existence of such physical systems with an event horizon
and thermodynamic properties similar to (3+1) dimensional black
holes. These vacuum solutions of (2+1)-dimensional gravity   are
asymptotically anti-de Sitter and are known as BTZ black holes. We
provide a new type of thin-shell stable wormhole  from the   BTZ
black holes.    This is the first example of stable thin shell
wormhole in (2+1)-dimension. Several characteristics of this
thin-shell wormhole have been discussed.

\end{abstract}

\keywords{Einstein's field equations; Stellar equilibrium.}

\maketitle

\section{Introduction}
\noindent Pure gravity in (2+1) dimensions has renewed interest in
recent years as gravity does not propagate out side the source, in
other words, matter curves spacetime only locally. Recently, very
interesting studies about $(2+1)$ gravastars, $(2+1)$ black holes,
$(2+1)$ wormholes  and $(2+1)$ dimensional stars have been
developed
\cite{FR2011,MB1993,CM1996,KC1996,KC1994,PM1996,SC1995,KC1997,RS2011,NC1995,rahaman2007e,Del1995,
Kim2004,EAL2007,Kim1993}.

 In recent past,
Morris and Thorne \cite{MT1988} have shown that general relativity
admits that the presence of matter twists the geometric fabric of
spacetime in such a way that  a bridge can be formed between two
asymptotic regions.  This bridge can be interpreted as instantons
describing a tunnelling between two distant regions. This
geometrical structure is known as Wormhole. However, according
Morris and Thorne \cite{MT1988}, the construction of wormhole
would require a very unusual form of matter, termed as exotic
matter that characterized the stress energy tensor.

The necessary ingredients that supply fuel to construct wormholes
remain an elusive goal for theoretical physicists. Several
proposals have been proposed   in literature
\cite{sus2005,lobo2005d,kuh1999,zas2005,rahaman2006f,rahaman2007d,lobo2005b,das2005,rahaman2006d,rahaman2009d,kuh2010,jamil2010,FR2007}.
Also some authors have used alternative theories of gravity to
exclude exotic matter, such as Brans-Dicke theory, Brain world,
C-field theory,
  Kalb-Ramond, Einstein-Maxwell theory etc
 \cite{Nandi1998,Cam2003,Lobo2007c,Rahaman2004,Rahaman2006a,Rahaman2009a,Rahaman2009b}.

  To minimize the usage of exotic matter, Visser \cite{Visser1989}    has proposed a way, which is known as
'Cut and Paste' technique,   to construct a wormhole in which the
exotic matter is concentrated at the wormhole throat.

Visser's tactics is more accessible  because it minimizes the
amount of exotic matter required.   For this reason, Visser's
approach was adopted by various authors to construct thin shell
wormholes
\cite{FR11,Poisson1995,Lobo2003,Lobo2004,Eiroa2004a,Eiroa2004b,Eiroa2005,
Thibeault2005,Lobo2005,Rahaman2006,
Eiroa2007,Rahaman2007a,Rahaman2007b,Rahaman2007c,
Lemos2007,Richarte2008,Rahaman2008a,
Rahaman2008b,Eiroa2008a,Eiroa2008b,Rahaman2010a, Rahaman2010b,
Rahaman2011,Dias2010,Peter2010}.

Recently, Ba\~{n}ados, Teitelboim and Zanelli (BTZ) \cite{BTZ1992}
has discovered an explicit vacuum solution of (2+1)-dimensional
gravity with negative cosmological constant. It has been argued
that the existence of such physical systems with an event horizon
and thermodynamic properties similar to (3+1) dimensional black
holes. These vacuum solution of (2+1)-dimensional gravity   are
asymptotically anti-de Sitter and are known as BTZ black holes.
Motivated by the properties of this black hole solution, the
authors have developed other interesting work about the geometry
of the spinning black holes of standard Einstein theory in $(2+1)$
dimension  \cite{MB1993}.

In this paper,  we present a new type of thin-shell wormhole
employing such a class of BTZ black holes by means of the
cut-and-paste technique  \cite{Visser1989}. As far as our
knowledge, this is the first example of stable thin shell wormhole
in (2+1)-dimension.  Various aspects of this thin-shell wormhole
are analyzed, particularly the equation of state relating pressure
and density. Also it has been discussed the attractive or
repulsive nature of the wormhole. We also make a survey  of
stability of this  wormhole.

\section{Thin-shell wormhole construction}
\noindent

 Ba\~{n}ados, Teitelboim and Zanelli (BTZ) \cite{BTZ1992}
have discovered an explicit vacuum solution of (2+1)-dimensional
gravity with negative cosmological constant as
\begin{equation}
ds ^2 = - \left(- M_0 - \Lambda r^2\right) dt^2 + \left(- M_0 -
\Lambda r^2\right)^{-1} dr^2 + r^2 d\theta^2.\label{eq22}
\end{equation}
The parameter $M_0$ is the conserved mass associated with
asymptotic invariance under time displacements. This mass is given
by a flux integral through a large circle at space-like infinity.

From the BTZ black hole, we can take two copies of the region with
$ r\geq a$ :

\[ M^\pm = ( x \mid r \geq a )  \]

and paste them at the  junction surface

\[ \Sigma = \Sigma^\pm = ( x \mid r = a )  \]

Here we take $ a > r_h $ to avoid horizon and this new
construction produces a geodesically complete manifold $ M = M^+
\bigcup M^- $ with a matter shell at the surface $ r = a $ , where
the throat of the wormhole is located.

Here the junction surface is a one dimensional ring of matter.
Let, $\eta$ denotes the Riemann normal coordinate at the junction.
We assume $\eta$  be positive and negative in two sides of the
junction surface.

Mathematically, we have $x^\mu = ( \tau,\theta,\eta) $  and the
normal vector components $\xi^\mu = ( 0,0,1)$ with the metric

\[  g_{\eta\eta} =1, ~~ g_{\eta i}~ =~0 ~~ and~~ g_{ij} = ~diag~ (~ -1,~r^2
~)~.
\]
\\

The second fundamental forms associated with the two sides of the
shell
\cite{Israel1966,Perry1992,Rahaman2006,Rahaman2010b,Rahaman2011}
are given by
\begin{equation}K^{i\pm}_j =  \frac{1}{2} g^{ik}
\frac{\partial g_{kj}}{\partial \eta}   \mid_{\eta =\pm 0} =
\frac{1}{2}   \frac{\partial r}{\partial \eta} \left|_{r=a}
 ~ g^{ik} \frac{\partial g_{kj}}{\partial r}\right|_{r=a}.
\label{eq36}
\end{equation}
So, the discontinuity in the second fundamental forms is given as
\begin{equation}\kappa _{ij} =   K^+_{ij}-K^-_{ij}.
\label{eq37}
 \end{equation}

Now, from Lanczos equation in (2+1) dimensional spacetime,   one
can obtain the surface stress energy tensor $ S_j^i = diag ( -
\sigma , -v) $      where, $\sigma$ and $v$ are line energy
density and line tension, respectively \cite{Perry1992} as

\begin{eqnarray} \sigma &=&  -\frac{1}{8\pi}  \kappa _\theta^\theta,\\
\label{eq38} v &=&  -\frac{1}{8\pi}  \kappa _\tau^\tau,
\label{eq39}
 \end{eqnarray}
To understand the dynamics of the wormhole, we assume  the radius
of the throat to be a function of proper time, or $ a = a(\tau)$.
Also, overdot and prime denote, respectively, the derivatives with
respect to $\tau$ and $a$.

 Employing relevant information into
Eqs. (1-5)   and setting $r=a$, we obtain
\begin{equation}\sigma =  -\frac{1}{4\pi a}  \left[ \sqrt{ K^2a^2 - M_0 +\dot{a}^2}
 \right],
\end{equation}
\begin{equation} p = -v =  \frac{1}{4\pi}  \left[ \frac{K^2 a + \ddot{a}}
{\sqrt{ K^2 a^2 - M_0 + \dot{a}^2}}  \right].
\end{equation}

In the above equations, the overdot   denotes,  the derivatives
with respect to $\tau$. Note that line tension ($v$) is negative
which implies that there is a line pressure (p) as opposed to a
line tension. Here, we have used $-\Lambda = K^2$.

For a static configuration of radius $a$, we obtain (assuming
$\dot{a} = 0 $ and $\ddot{a}= 0 $) from  Eqs. (6) and (7)

\begin{equation}\sigma =  -\frac{1}{4\pi a}  \left[ \sqrt{ K^2a^2 - M_0  }
 \right],
\end{equation}
\begin{equation} p = -v =  \frac{1}{4\pi}  \left[ \frac{K^2 a  }
{\sqrt{ K^2 a^2 - M_0  }}  \right].
\end{equation}

Observe that the energy density $\sigma$ is negative, however, the
pressure $p$ is positive.   Moreover, on this shell, which is
infinitely thin, the radial pressure is zero. One can note that
  $p + \sigma$, $ \sigma + 2p  $  are positive. So the
shell contains matter that violates only the null energy condition
(NEC) and obeys the weak and strong energy condition. For the
assumed case, variations
 of left hand side of the expressions of energy conditions have been shown in
Figure 1.

\begin{figure}[ptb]
\begin{center}
\vspace{0.2cm}
\includegraphics[width=0.6\textwidth]{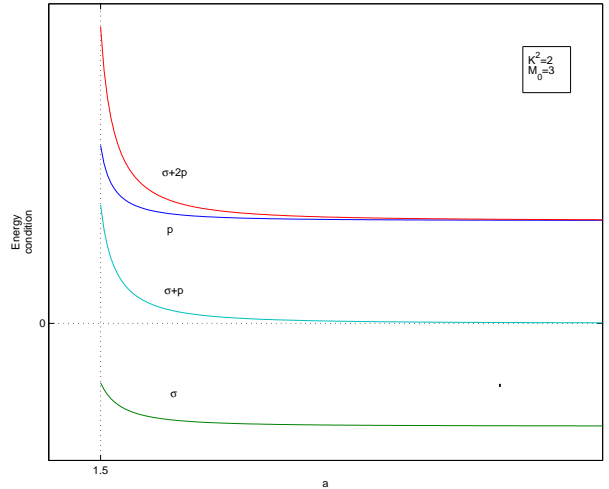}
\end{center}
\caption{Variations of   the expressions of energy conditions
shown against $a$.} \label{fig5}
\end{figure}

\section{The gravitational field}

\noindent Now, to study the nature of the gravitational field of
the wormhole constructed from BTZ black hole, we calculate the
observer's three-acceleration $a^\mu = u^\mu_{\,\,;\nu} u^\nu$,
where $u^{\nu} =  d x^{\nu}/d {\tau} =(1/\sqrt{K^2 r^2 - M_0)},
0,0)$. In view of the line element, Eq. (1), the only non-zero
component is given by

\begin{equation}
a^r = \Gamma^r_{tt} \left(\frac{dt}{d\tau}\right)^2 \\= K^2 r
\end{equation}
A radially moving test particle initially at rest obeys the
equation of motion
\begin{equation}
\frac{d^2r}{d\tau^2}= -\Gamma^r_{tt}\left(\frac{dt}{d\tau}
\right)^2 =-a^r.
\end{equation}
If $a^r=0$, we obtain the geodesic equation.

However, we note that the wormhole is attractive since $a^r>0$.
This result is similar to the case of the thin shell wormhole
constructed from Schwarzschild black hole\cite{Rahaman2010a}.

\section{ Time evolution of radius of the throat: }

  Equations (8) and (9) can be written in the form
\begin{equation}
              v_0 =  w(a)  \sigma_0
               \end{equation}
where
\begin{equation}
              w(a) = \frac{K^2a^2}{K^2a^2-M_0}
               \end{equation}

               [ suffix 0 indicates the static situation ]

Following Eiroa et al \cite{Eiroa2004b}, we assume that the
equation of state does not depend on the derivative of $a(\tau)$
i.e. it is the same form as in the static one. Now putting $\sigma
$, v in place of $\sigma_0$ , $v_0$ from (7) and (8) in (12), we
get the following expression as
\begin{equation}
               \end{equation}
This implies,
\begin{equation}
             \dot{a}^2(\tau) = \dot{a}^2(\tau_0) \left[\frac{K^2a^2(\tau_0) -M_0}{K^2 a^2(\tau) -M_0} \right]
               \end{equation}
               Here, $\tau_0$ is arbitrary fixed time.

After integration, this gives,
   \begin{multline}
a(\tau)\sqrt{K^2 a^2(\tau) -M_0}-a(\tau_0)\sqrt{K^2 a^2(\tau_0) -M_0}\\
-\frac{M_0}{K} \left[ \cosh^{-1}\left(
\frac{a(\tau)K}{\sqrt{M_0}}\right) - \cosh^{-1}\left(
\frac{a(\tau_0)K}{\sqrt{M_0}}\right)\right] \\= 2\dot{a}(\tau_0)
\sqrt{ K^2a^2(\tau_0) -M_0} (\tau -\tau_0)
                \end{multline}

 The above implicit expression gives the time evolution
of the radius of the throat.

 The velocity and acceleration of the throat are
\begin{equation}
             \dot{a}(\tau) =\dot{a}(\tau_0) \sqrt{\left[\frac{K^2a^2(\tau_0) -M_0}{K^2 a^2(\tau) -M_0} \right]}
               \end{equation}
               and
\begin{equation}
             \ddot{a}(\tau) = - \dot{a}^2(\tau_0)K^2 a(\tau) \left[\frac{K^2a^2(\tau_0) -M_0}{\left\{K^2 a^2(\tau) -M_0\right\}^2} \right]
                 \end{equation}

From the above two expressions indicate that sign of the initial
velocity determines  sign of the velocity and   the acceleration
is always negative. It is immaterial whether the initial velocity
is positive or negative, the throat expands forever. This would
imply that the equilibrium position is always unstable. However,
if the initial velocity is zero, the velocity and acceleration of
the throat would be zero i.e. throat be in static equilibrium
position.

\section{The total amount of exotic matter}
\noindent Now, we determine the total amount of exotic matter
confined within the shell.  This total amount of exotic matter can
be quantified by the integral
 \cite{Eiroa2005,Thibeault2005,Lobo2005,Rahaman2006,
Eiroa2007,Rahaman2007a,Rahaman2007b}
\begin{equation}
   \Omega_{\sigma}=\int [\rho+p_r]\sqrt{-g}d^2x.
\end{equation}
By introducing the radial coordinate $R=r-a$, w get
\[
 \Omega_{\sigma}=\int^{2\pi}_0 \int^{\infty}_{-\infty}
     [\rho+p_r]\sqrt{-g}\,dR d\theta.
\]
Since the shell is infinitely thin, it does not exert any radial
pressure.  Moreover, energy density is located on the thin shell
surface, $\rho=\delta(R)\sigma(a)$. Then we have,
\begin{multline}\label{E:amount}
 \Omega_{\sigma}=\int^{2\pi}_0 \left.[\rho\sqrt{-g}]
   \right|_{r=a} d\theta=2\pi a\sigma(a)\\
=- \frac{1}{2}\sqrt{K^2a^2-M_0}
\end{multline}

 This NEC violating matter can be
reduced by taking the value of $a$ closer to $r_h$, the location
of the   event horizon. The closer $a$ is to $r_h$, however, the
closer the wormhole is to a black hole: incoming microwave
background radiation would get blueshifted to an extremely high
temperature \cite{tR93}. The variation of the total amount of
exotic matter with respect to the conserved mass and cosmological
constant   can best be seen graphically (Figure 2). Observe that
the NEC violating matter  on the thin shell can be reduced by
  increasing the conserved mass.

\begin{figure}
\begin{center}
\vspace{0.2cm} \includegraphics[width=0.55\textwidth]{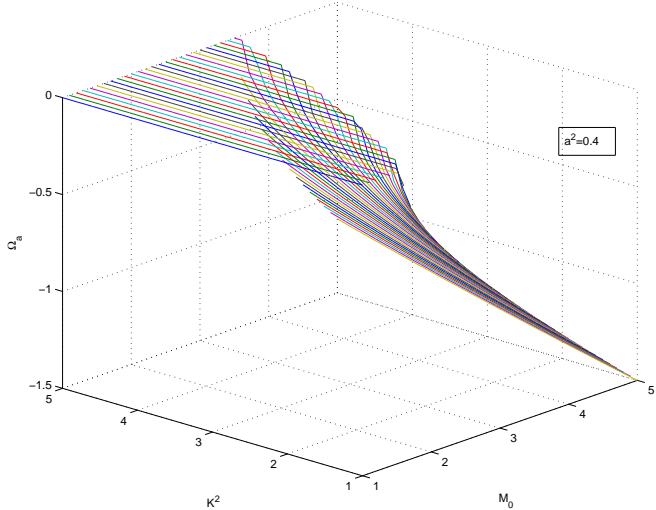}
\caption{The variation in the total amount of exotic matter on the
shell with respect to the mass ($ M_0 $) and  ($ K^2 $) for
$a^2=0.4$.} \label{fig:3}
\end{center}
\end{figure}

\section{Stability}

Recall, if the initial velocity is zero, the velocity and
acceleration of the throat would be zero i.e. throat be in static
equilibrium position. Now, we shall study the stability of the
configuration   using two different approaches: (i) assuming
Chaplygin gas   equation of state on the thin shell and (ii)
analyzing the stability to linearized radial perturbations around
static solution situated at $a_0$.

\subsection{Chaplygin gas equation
of state}
  We analyze the stability of the shell taking Chaplygin gas equation
of state  at the throat.

We re-introduce an equation of state between the surface pressure
$p$ and surface energy density $\sigma$ as
\begin{equation} p=- \frac{A}{\sigma^\alpha}   \end{equation}
 with  $  A  $ and $\alpha $ are positive constants.

Putting equations (6) and (7) in equation (21), one gets the
differential equation for the throat radius of thin shell
wormholes threaded by Chaplygin gas-like exotic matter is
\begin{equation}\frac{(-1)^\alpha}{(4\pi)^{\alpha+1}a^\alpha}(K^2a+\ddot{a}) (K^2a^2-M_0+\dot{a}^2)^\frac{\alpha-1}{2}+A =0   \end{equation}
  If the
static solution exists (i.e. $\dot{a} = 0$, $ \ddot{a} = 0$), then
the throat radius ($a = a_0$) should satisfy the equation given by

\begin{equation}  \frac{(-1)^\alpha  K^2a}{(4\pi)^{\alpha+1}a^\alpha} (K^2a^2-M_0)^\frac{\alpha-1}{2}+A   = 0  \end{equation}

Solution of equation (23) gives the radius of the throat for $
\alpha < 1$. Assuming LHS of equation (23) as $H(a) =0$  and  we
plot $H(a)$ vs. $a$. $H(a)$ cuts $ a $ axis at \
 some $a =a_0$, gives the radius of the throat ( see figure 3 ).  From
figure 3, one can observe  that radius of the throat is
increasing with decreasing    $\alpha$.  We should exclude the
case $\alpha = 1$ as for this value, equation (23) is
inconsistent.  However, if we relax the restriction on  $\alpha
>1$, one can also get the radius of the throat of the thin-shell
wormholes constructed from the BTZ black holes  ( see figure 4 ).
\begin{figure}
\begin{center}
\vspace{0.2cm} \includegraphics[width=0.6\textwidth]{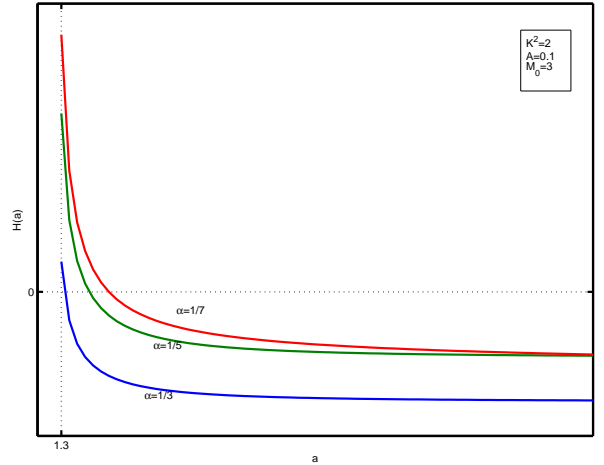}
\caption{Assuming LHS of equation (23) as $H(a) =0$  and  we plot
$H(a)$ vs. $a$. $H(a)$ cuts $a$ axis at \
 some $a =a_0$, gives the radius of the throat. For different
values of  $\alpha$, we get different values of radius of the
throat. } \label{fig:3}
\end{center}
\end{figure}
\begin{figure}
\begin{center}
\vspace{0.2cm} \includegraphics[width=0.6\textwidth]{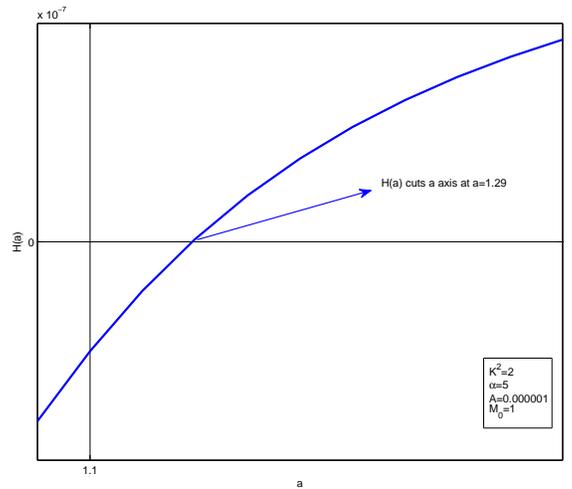}
\caption{  $H(a)$ cuts $a$ axis at \
 some $a =a_0$ for $\alpha>1$. } \label{fig:3}
\end{center}
\end{figure}

  Here $ p$ and
$\sigma $ obey the conservation equation

\begin{equation}
               \dot{\sigma} +    \frac{\dot{a}}{a}( p + \sigma ) = 0.
               \end{equation}
In the above equations, the overdot   denotes,   the derivative
with respect to $\tau$.

Using (21), from (24), we get
\begin{equation}
               \dot{\sigma} +    \frac{\dot{a}}{a}\left( \frac{\sigma^{\alpha+1}-A}{\sigma^\alpha}  \right) = 0.
               \end{equation}
Solution of the above equation can be found as
\begin{equation}
              \sigma =\left[A+\left(\sigma_0^{\alpha+1}-A\right)\left(\frac{a_0}{a}\right)^{\alpha+1}\right]^\frac{1}{\alpha+1}
               \end{equation}
               where, $\sigma_0 = \sigma (a_0)$.\\

Equation (6) implies
\begin{equation}
\dot{a}^2 + V(a)= 0.
\end{equation}
Here the potential $V(a)$ is defined  as
\begin{equation}
V(a) =K^2
a^2-M_0-16\pi^2a^2\left[A+\left(\sigma_0^{\alpha+1}-A\right)\left(\frac{a_0}{a}\right)^{\alpha+1}\right]^\frac{2}{\alpha+1}
\end{equation}
Expanding $V(a)$ around $a_0$, we obtain
\begin{eqnarray}
V(a) &=&  V(a_0) + V^\prime(a_0) ( a - a_0) +
\frac{1}{2} V^{\prime\prime}(a_0) ( a - a_0)^2  \nonumber \\
&\;& + O\left[( a - a_0)^3\right],
\end{eqnarray}
where the prime denotes the derivative with respect to $a$. Since
we are linearizing around $ a = a_0 $, we require that $ V(a_0) =
0 $ and $ V^\prime(a_0)= 0 $.  The configuration will be in stable
equilibrium if $ V^{\prime\prime}(a_0)> 0 $.  Using the
conditions, $ V(a_0) = 0 $ and $ V^\prime(a_0)= 0 $, we obtain
from equation (28) as
\begin{equation}V^{\prime\prime}(a_0) = \frac{2(\alpha-1) M_0 K^2 }{K^2a_0^2-M_0}  \end{equation}
We find that the inequality $ V^{\prime\prime}(a_0)> 0 $ can only
be satisfied if  $\alpha > 1$. So we conclude that   the wormhole
is stable. However, for $\alpha < 1$ , $ V^{\prime\prime}(a_0)< 0
$  and the wormhole is unstable.\\
 One can also analysis the stability by means of the figures.\\ The plot (figure 5)
indicates that $V(a)$ has a local minimum at some $a$. In other
words, it is stable for $\alpha
> 1$. However, the plot (figure 6) indicates that
$V(a)$ has a local maximum at some $a$. In other words, it is
unstable for $\alpha < 1$.  Thus for some specific choices of the
parameter   $\alpha $, the thin-shell wormholes constructed from
the  BTZ black holes is stable.
\begin{figure}
\begin{center}
\vspace{0.5cm}
\includegraphics[width=0.65\textwidth]{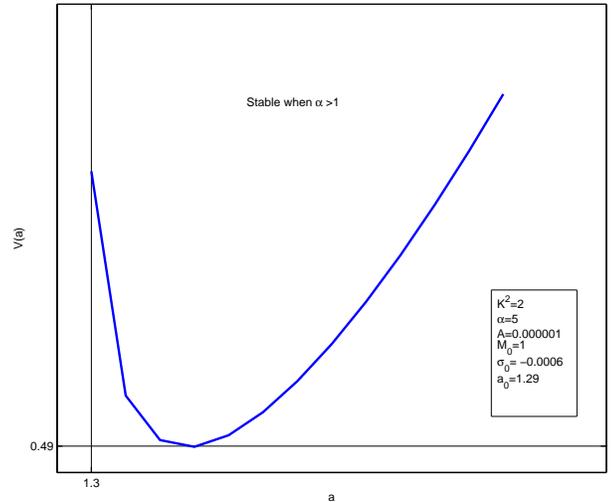}
\caption{ The variation of $V(a)$  with respect to $a$ for $\alpha
> 1$. The wormhole is stable.}
        \label{fig17}
\end{center}
\end{figure}
\begin{figure}
\begin{center}
\vspace{0.5cm}
\includegraphics[width=0.6\textwidth]{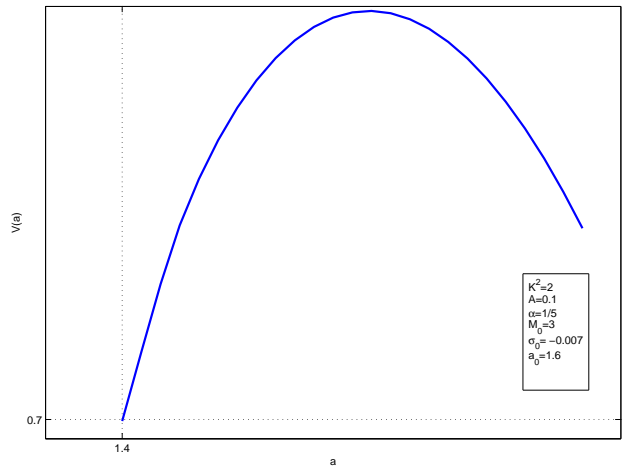}
\caption{The variation of $V(a)$  with respect to $a$ for $\alpha
< 1$. The wormhole is unstable.}
        \label{fig17}
\end{center}
\end{figure}

\subsection{Linearized radial
perturbations}

From equation (27), the potential $V(a)$ can be written in terms
of $\sigma$ as
\begin{equation}
       V(a)      = K^2 a^2 -M_0 - 16 \pi^2 a^2 \sigma^2.
               \end{equation}
As above, expanding $V(a)$ around $a_0$, we obtain
\begin{eqnarray}
V(a) &=&  V(a_0) + V^\prime(a_0) ( a - a_0) +
\frac{1}{2} V^{\prime\prime}(a_0) ( a - a_0)^2  \nonumber \\
&\;& + O\left[( a - a_0)^3\right],
\end{eqnarray}
where the prime denotes the derivative with respect to $a$. Since
we are linearizing around $ a = a_0 $, we require that $ V(a_0) =
0 $ and $ V^\prime(a_0)= 0 $.  The configuration will be in stable
equilibrium if $ V^{\prime\prime}(a_0)> 0 $. The subsequent
analysis will depend on a parameter $\beta$, which is usually
interpreted as the subluminal speed of sound and is given by the
relation
\[
 \beta^2(\sigma) =\left. \frac{ \partial
p}{\partial \sigma}\right\vert_\sigma.
\]
Recall conservation equation (24) which readily yields,
\begin{equation}
       \sigma^{\prime\prime}-\frac{1}{a^2}(p+\sigma)(2+\beta^2)=0.
               \end{equation}

Differentiating twice the potential $V(a)$, we get,
\begin{equation}
       V^{\prime\prime}(a) = 2K^2-32\pi^2\beta^2(p\sigma+\sigma^2)-32\pi^2p^2 .
               \end{equation}
Replacing $\sigma^{\prime\prime}$ from equation (31), we find
\begin{equation}
        V^{\prime\prime}(a) =2K^2+\frac{2}{a^2}M_0\beta^2-\frac{2K^4a^2}{K^2 a^2-M_0}
               \end{equation}
Now, for stability $  V(a_0)^{\prime\prime} >0$ implies
\begin{equation}
      {\beta_0}^2 < \frac{K^2 {a_0}^2}{K^2{a_0}^2-M_0}
               \end{equation}

There is a region of stability corresponding to
$\beta_0^2<\frac{K^2 {a_0}^2}{K^2{a_0}^2-M_0}$.
 Figure 7 shows a typical region of stability ( below the curve ) by
choosing $M_0=3$ and $K^2 =2$.   However, if the radius of the
throat is large enough, then the region of stability barely
reaches down to $\beta_0^2=1$.

It is known that for ordinary matter, $\beta^2$ represents the
velocity of sound,  however, according to Poisson et al
\cite{Poisson1995}   this interpretation of $\beta^2$ can be
questioned as we are dealing with exotic matter and wormhole
configurations with $\beta_0^2>1$ should not be ruled out.
Although the  small value of the radius of the throat $\beta_0^2$
  greater than one, however, for large value of that,
$\beta_0^2 \approx 1$. This implies the region of stability lies
in  region for which stable wormholes might   be physically
acceptable.

\begin{figure}
\begin{center}
\vspace{0.5cm}
\includegraphics[width=0.6\textwidth]{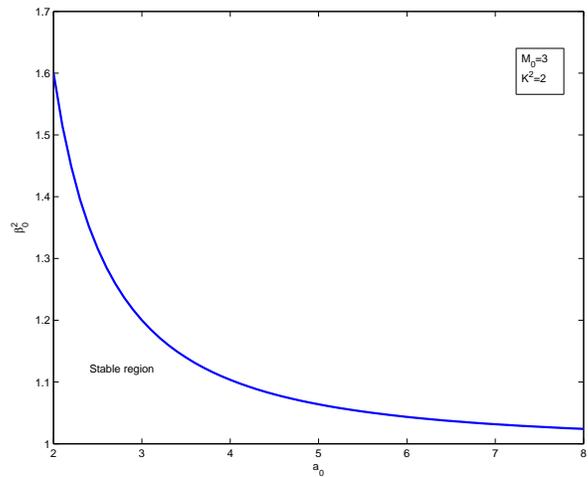}
\caption{The region of stability is below the curve.}
        \label{fig17}
\end{center}
\end{figure}

\section{Final remarks}\noindent
In this paper, a new type of thin-shell wormhole has been
developed using the BTZ black holes and the cut-and-paste
technique. We provide an analisys of some aspects of the
thin-shell wormhole pointing out the equation of state relating
pressure and density, the attractive or repulsive nature of the
gravitational field of the wormhole and its stability.

We obtained that the energy density $\sigma $ is negative and the pressure $%
p $ is positive. We get,  $p+\sigma >0$ and $\sigma +2p>0$, and we
can conclude that the matter contained by the shell violates only
null energy condition (NEC), but obeys the weak and strong energy
conditions.

From the study of the gravitational field of the wormhole,  we
obtain that the wormhole is attractive since $a^{r}>0$. We notice
the similarity of this result with the one obtained in the case of
the thin shell wormhole constructed from Schwarzschild black hole
\cite{Rahaman2010a}.

The stability of the wormhole is studied using two particular
cases: (i) assuming Chaplygin gas equation of state on the thin
shell and (ii) making an analisys of the stability to linearized
radial perturbations around static solution situated at $a_{0}$.
In the first case, we conclude that the thin wormhole which is
constructed from BTZ black hole threaded by Chaplygin gas equation
of state is stable. In the second case, of linearized radial
perturbations, there is a region of stability corresponding to
$\beta_0^2<\frac{K^2 {a_0}^2}{K^2{a_0}^2-M_0}$.

It is known that for ordinary matter, $\beta^2$ represents the
velocity of sound,  however, according to Poisson et al
\cite{Poisson1995}   this interpretation of $\beta^2$ can be
questioned as we are dealing with exotic matter and wormhole
configurations with $\beta_0^2>1$ should not be ruled out.
Although the  small value of the radius of the throat $\beta_0^2$
  greater than one, however, for large value of that,
$\beta_0^2 \approx 1$. This implies the region of stability lies
in  region for which stable wormholes might   be physically
acceptable.

\subsection*{Acknowledgments}

 FR is thankful to Inter-University Centre for
Astronomy and Astrophysics, Pune, India for providing Visiting
Associateship under which a part of this work is carried out. FR
is also grateful to PURSE, Govt. of India
 for financial
support.

\end{document}